\documentclass[aps,pra,twocolumn,groupedaddress]{revtex4-1}

\pdfoutput=1

\usepackage{eurosym}
\usepackage{amsfonts}
\usepackage{amsmath}
\usepackage{amssymb,epsf}
\usepackage{color}
\usepackage{epstopdf}
\usepackage{graphicx}
\usepackage{float}
\usepackage{caption}
\usepackage{subfig}
\usepackage{hyperref}
\usepackage{mathtools}
\usepackage{lipsum}

\definecolor{aogreen}{rgb}{0.0, 0.5, 0.0}

\def\ketm#1{  \left\vert  #1   \right\rangle   }

\def\bram#1{  \left\langle  #1   \right\vert   }

\begin{document}

\title{Simulation of the multiphase configuration and phase transitions with quantum walks utilizing a step-dependent coin}

\author{S. Panahiyan $^{1,2,3}$}
\email{email address: shahram.panahiyan@uni-jena.de}
\author{S. Fritzsche $^{1,2,3}$ }
\email{email address: s.fritzsche@gsi.de}
\affiliation{$^1$Helmholtz-Institut Jena, Fr\"{o}belstieg 3, D-07743 Jena, Germany  \\
	$^2$GSI Helmholtzzentrum f\"{u}r Schwerionenforschung, D-64291 Darmstadt, Germany \\
	$^3$Theoretisch-Physikalisches Institut, Friedrich-Schiller-University Jena, D-07743 Jena, Germany}

\date{\today}

\begin{abstract}
Quantum walks are versatile simulators of topological phases and phase transitions as observed in condensed matter physics. Here, we utilize a step dependent coin in quantum walks and investigate what topological phases we can simulate with it, their topological invariants, bound states and possibility of phase transitions. These quantum walks simulate non-trivial phases characterized by topological invariants (winding number) $\pm 1$ which are similar to the ones observed in topological insulators and polyacetylene. We confirm that the number of phases and their corresponding bound states increase step dependently. In contrast, the size of topological phase and distance between two bound states are decreasing functions of steps resulting into formation of multiple phases as quantum walks proceed (multiphase configuration). We show that, in the bound states, the winding number and group velocity are ill-defined, and the second moment of the probability density distribution in position space undergoes an abrupt change. Therefore, there are phase transitions taking place over the bound states and between two topological phases with different winding numbers. 
\end{abstract}

\maketitle

\section{Introduction}
In contrast to symmetry-breaking phases such as ferromagnetic and superconducting phases, topological phases are states of matter that are symmetry-preserving \cite{Thouless}. The topological phases do not exhibit any local order parameters and can not be described by them \cite{Kitaev,Fidkowski}. In fact, they are described by global topological orders known as topological invariants which characterize the global structure of ground-state wave function. The presence of global order is the reason for phenomena such as fractional charges and magnetic monopoles \cite{Qi2008,Hasan2010}, integer Hall effect \cite{Thouless} and existence of topological insulators \cite{Kane,Bernevig,Fu,Koenig,Hsieh}. 

Quantum walks have been found universal frameworks \cite{Lovett} in quantum information and computation which can be applied to simulate other quantum systems and phenomena such as photosynthetic energy transfer \cite{Mohseni} and slow dynamics in a nonlinear and disordered medium \cite{Vakulchyk}. 

More recently, it was shown that the quantum walks are also versatile simulators of topological phenomena that are observed in condensed-matter physics \cite{Kitagawa}. In particular, it was proven that they can realize all known kinds of topological phases in one and two dimensions \cite{Kitagawa,Kitagawa2012,Asboth,Obuse,Chen}. The bulk-boundary correspondence was investigated in Refs. \cite{Asboth2013,Tarasinski}. The possibility of topological phase transitions is addressed in Refs. \cite{Rakovszky,Mera}. It was shown that quantum walks are suitable for direct extraction of topological invariants \cite{Ramasesh}. Photonic quantum walks were also explored experimentally in order to observe topologically protected bound states \cite{KitagawaExp} and topological quantum transitions \cite{Cardano,Wang}, to simulate Zak phases \cite{Cardano2017} or to detect topological invariants \cite{Barkhofen,Flurin,Zhan,Xiao}. The controllability over the walker's behavior in the quantum walks allows one to check the robustness of the
bound states \cite{KitagawaExp} and helps suppress the restriction in studying the dynamics of strongly-driven systems. 

In this paper, we utilize a step-dependent coin in the quantum walks \cite{Panahiyan}. We show that these quantum walks have chiral symmetry with two dynamical energy bands. This results in formation of multiple phases with different topological invariants (without split step quantum walk modification \cite{Kitagawa,Kitagawa2012,Asboth,Obuse,Chen}). The resultant topological phases are non-trivial ones similar to the phases observed in topological insulators and polyacetylene. Contrary to previous studies, in this work, we show that the size of topological phases and, consequently, the number of topological phases and bound states are step dependent. We also prove that group velocity and topological invariants become ill-defined in a bound state which signals quantum phase transition. Finally, we confirm that the behavior of group velocity and energy around each bound state enables us to determine topological invariants without calculating them. 

The structure of the paper is as follows. First, we calculate the energy and discuss its behavior as a function of steps, momentum and coin's parameter in Sec. \ref{EnergyI}. Next, we find topological invariants in Sec. \ref{TopologyI}, group velocity and second moment for different topological phases and investigate the possibility of phase transitions through them in Sec. \ref{GroupI}. The details of calculations and justifications for different claims are provided in the Appendix \ref{Append}.

\section{Energy, group velocity and topological phases}

\begin{figure*}[htb]
	\centering
		{\begin{tabular}[b]{cc}%
			\subfloat[\label{SSH}]{\includegraphics[width=0.6\linewidth]{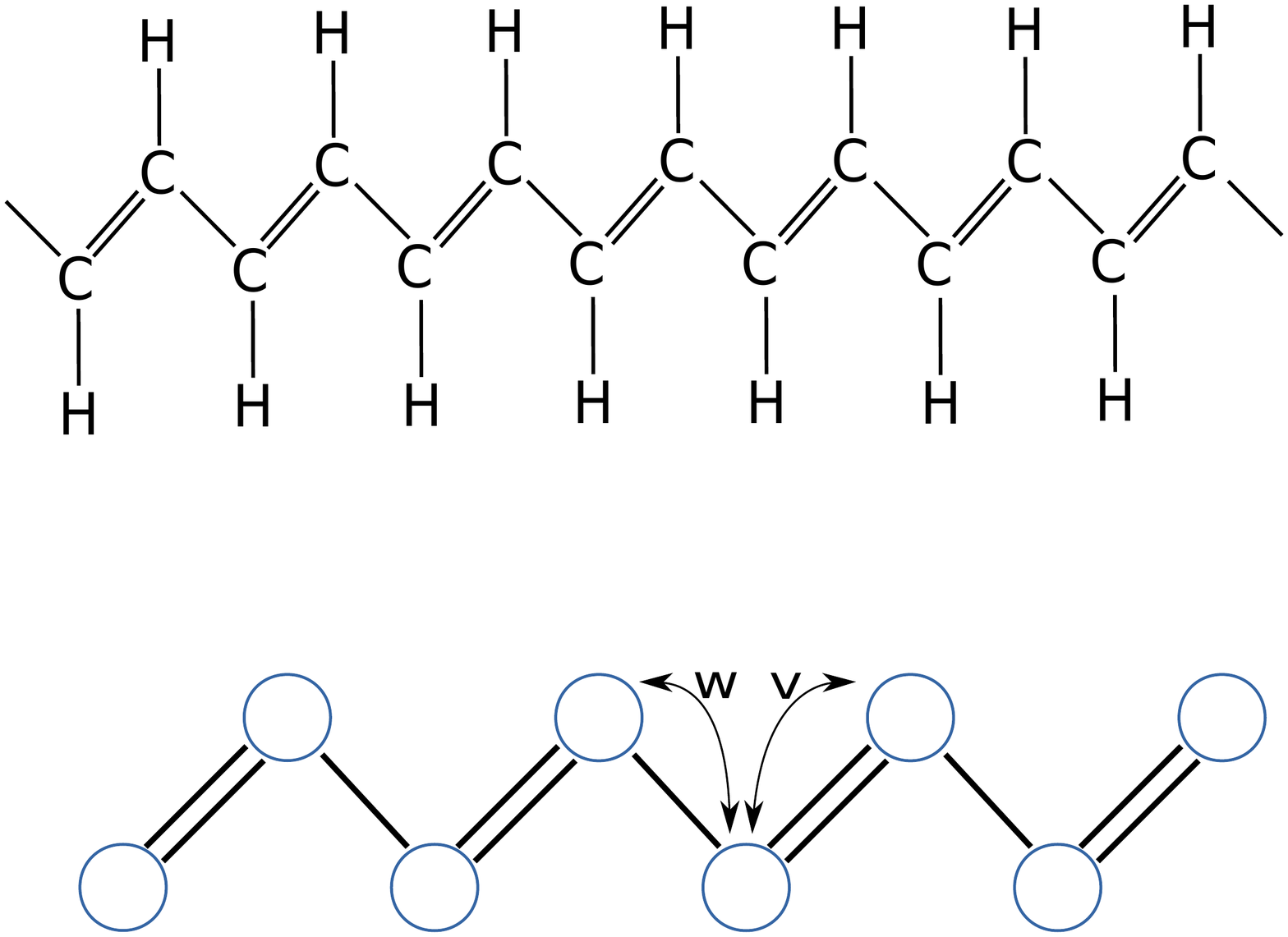}}
			{\begin{tabular}[b]{c}%
					\subfloat[\label{QW}]{\includegraphics[width=0.35\linewidth]{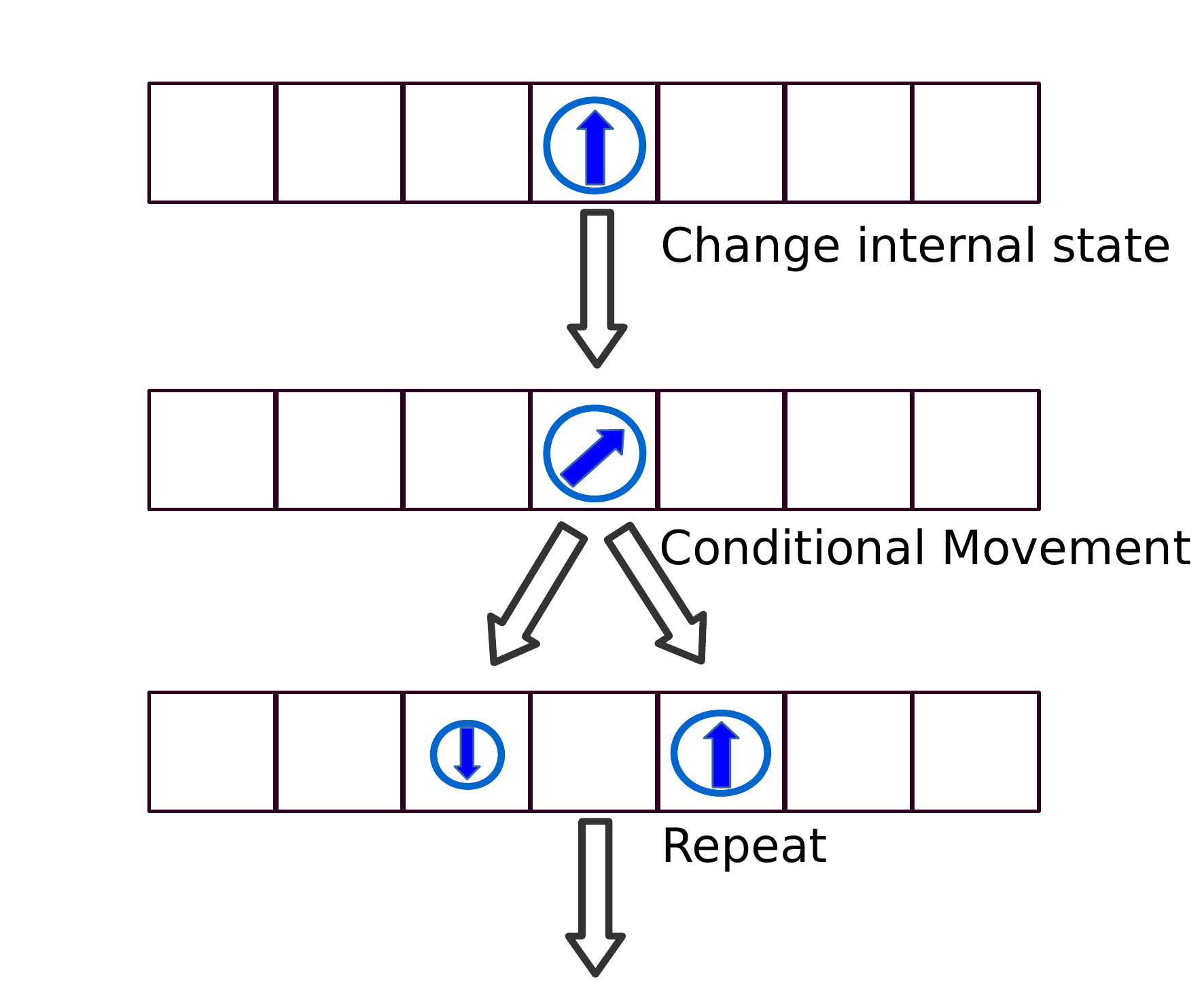}}\\
					\subfloat[\label{CFG}]{\includegraphics[width=0.36\linewidth]{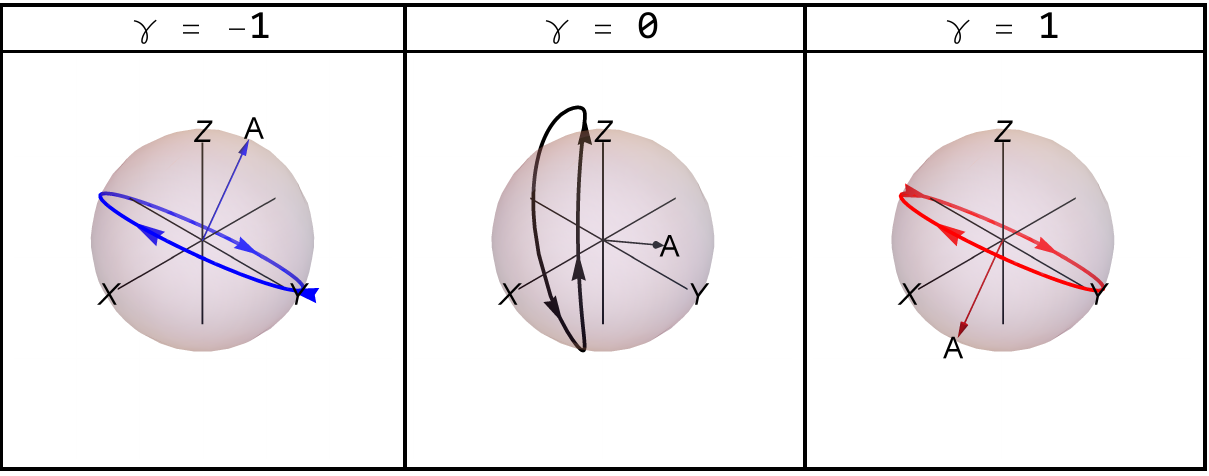}}
			\end{tabular}}					
		\end{tabular}}						
	\caption{a) Chain of polyacetylene molecules in which electrons hope from one carbon to a neighboring carbon and corresponding geometry of the Su-Schrieffer-Heeger model describing electrons hopping on a chain (carbon) with staggered hopping amplitudes of $v$ and $w$. b) One step of a quantum walk in which first the internal state of the walker is changed by coin operator and then the walker is moved by one unit position to the right (left) if its internal state is up (down) by a shift operator. c) Topological invariant (winding number) for three different phases that are observed in polyacetylene with the SSH model. The closed loops are $\boldsymbol n$ as $k$ traverses the first Brillouin zone around the origin in a plane orthogonal to $\boldsymbol A$.} \label{Fig0}
\end{figure*}

The quantum walks have a protocol which includes a shift-coin operator that is applied successively on a walker (a particle). The coin operator acts on the internal state of the walker and determines the relative probability amplitudes of the subsequent movement of the walker. The shift operator moves the walker to two nearest-neighbor positions based on its internal state (Fig. \ref{Fig0}). 

The quantum walks have significant similarity to the Su-Schrieffer-Heeger (SSH) model which describes electrons (or spinless fermions) hopping on a one-dimensional lattice, with staggered hopping amplitudes \cite{Su,Asboth2016} (Fig. \ref{Fig0}). Therefore, it was proposed and shown that quantum walks can be used as simulators for topological phases investigated in condensed-matter physics \cite{Kitagawa,Kitagawa2012}. 

Previously, it was pointed out that, instead of the usual protocol of quantum walks (shift-coin operator), one should use the protocol of split step quantum walks to simulate different topological phases in one and two dimensions, and increase the number of topological phases (multi phase configuration) \cite{Kitagawa,Kitagawa2012,Asboth,Obuse,Chen}. The protocol of the split-step quantum walks includes four different coin and shift operators (shift-coin-shift-coin operator). Here, in contrast, we show that by a step-dependent coin and without split step protocol, quantum walks simulate different topological phases and multi phase configuration. 

\subsection{Hamiltonian and Energy} \label{EnergyI}

The quantum walk is done by $T$ times successive application of the shift-coin operator on an initial state of a walker:

\begin{eqnarray}
\ketm{\phi}_{fin} & = & \ketm{\phi}_{T}
\;=\; \widehat{U}^{\:T}\: \ketm{\phi}_{int}\;=\; (\widehat{S}\widehat{C})^{\:T}\: \ketm{\phi}_{int} \, \label{protocol}.
\end{eqnarray}

For quantum walks with a step-dependent coin and two internal states \cite{Panahiyan}, we consider the coin operator to be

\begin{eqnarray*}
\widehat{C} & = & \cos (\frac{T\theta}{2})\: \ketm{0}_{C} \bram{0} \:-\:
\sin (\frac{T\theta}{2})\: \ketm{0}_{C} \bram{1}                              \notag \\[0.1cm]
\label{coin}
&   & \quad +\:
\sin (\frac{T\theta}{2})\: \ketm{1}_{C} \bram{0} \:+\:
\cos (\frac{T\theta}{2})\: \ketm{1}_{C} \bram{1} \:= e^{-\frac{i T \theta}{2}\sigma_{y}}
\end{eqnarray*}
where $T$ characterizes the step dependency of the coin operator, $\theta$ can vary throughout $[0,2 \pi]$ and $\sigma_{y}$ is the Pauli matrix. The conditional shift operator in one-dimensional position space is 

\begin{eqnarray}
\widehat{S} & = & \ketm{0}_{C} \bram{0} \otimes \sum_{x} \ketm{x+1}_{P} \bram{x}            \notag \\[0.1cm]
&   & \quad +\:
\ketm{1}_{C} \bram{1} \otimes \sum_{x} \ketm{x-1}_{P} \bram{x}\, ,
\end{eqnarray}
which, by using the discrete Fourier transformation, we can rewrite as \cite{Kitagawa} 

\begin{equation}
\widehat{S}= \sum_{k} (e^{i k}\ketm{0}_{C} \bram{0} + e^{-i k}\ \ketm{1}_{C} \bram{1}) \otimes\, \ketm{k} \bram{k}=e^{ik \sigma_{z}}. \label{shift}
\end{equation}

Since quantum walks are the result of repeated application of a unitary (coin-shift) operator (kind of a time-periodic driving system), it can be described in the framework of Floquet theory \cite{Kitagawa,Kitagawa2012,Asboth,Obuse,Chen}. Consequently, one can map the unitary evolution of the protocol of quantum walks to stroboscopic evolution under an effective Hamiltonian and use quantum walks to simulate topological phases described by the effective Hamiltonian. The effective Hamiltonian is associated with a full period which is diagonal in momentum space and can be written as \cite{Cardano,Cardano2017}

\begin{equation}
\widehat{H}(k)=i \ln\widehat{U}(k)= E(k) \boldsymbol n(k)\cdot \boldsymbol \sigma, \label{Hamiltonian}
\end{equation}
where $E(k)$ is the quasi-energy dispersion, $\boldsymbol \sigma$ is Pauli matrices, and $\boldsymbol n(k)$ defines the quantization axis for the coin eigenstates at each quasi-momentum $k$. In what follows, we use energy and momentum instead of quasi-energy and quasi-momentum, respectively. Due to the lattice translation symmetry of $\widehat{U}$ and two internal states for the walker, we can find two bands of energy parameterized by momentum as (see the Appendix for more details)

\begin{equation}
E(k)= \pm\cos^{-1} \bigg[ \cos(\frac{T \theta}{2}) \cos(k) \bigg]. \label{energy}
\end{equation}

It is a matter of calculation to find $\boldsymbol n(k)$ as

\begin{eqnarray*}
\boldsymbol n(k) =\frac{\left(\cos (\frac{T\theta}{2}) \sin (k),\sin (\frac{T\theta}{2}) \cos (k),- \cos (\frac{T\theta}{2}) \sin (k) \right)}{\sin E(k)}. \label{n}
\end{eqnarray*} 

Our quantum walks in this paper features chiral symmetry since there is an operator, $\widehat{\Gamma}$, which satisfies two conditions of $\widehat{\Gamma}^2=I$ and $\widehat{\Gamma}\widehat{H}\widehat{\Gamma}=-\widehat{H}$ (for more details, see Refs. \cite{Cardano,Asboth2016}). To find this operator, we first introduce $\boldsymbol A$, which is a vector labeling a point on the Bloch sphere and it is perpendicular to $\boldsymbol n(k)$ for all $k$: 

\begin{eqnarray}
\boldsymbol A=\left(\cos (\frac{T\theta}{2}),0, \sin (\frac{T\theta}{2})\right).
\end{eqnarray}
Then, we find $\widehat{\Gamma} =\boldsymbol A \cdot \boldsymbol\sigma$ that satisfies both of the conditions pointed out before (see appendix for more details). 

The two bands of energy are limited within the interval of $[-\pi,\pi]$ and the energy has symmetry of $E(k)=E(-k)$. The gap between these two bands of energy could close up and edge (bound) states are formed. The bulk states (topological phases) are located between two bound states. Therefore, the bound states separate different phases from one another and possible phase transitions take place over them and between two phases with different topological invariants.

\begin{figure*}[htb]
	\centering
	{\begin{tabular}[b]{cc}%
		\subfloat[\label{G1}]{\includegraphics[width=0.4\linewidth]{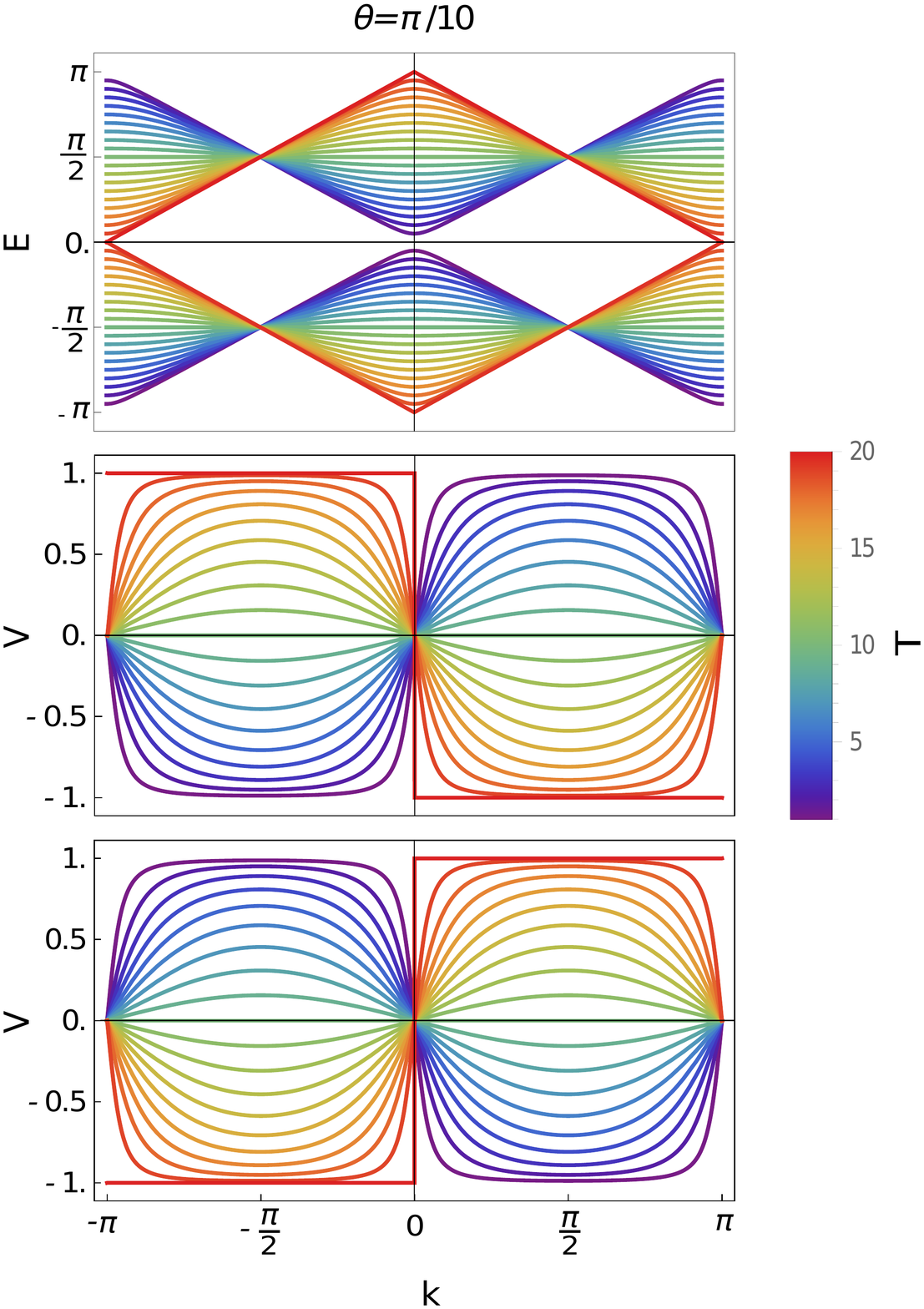}}
		\subfloat[\label{G2}]{\includegraphics[width=0.4\linewidth]{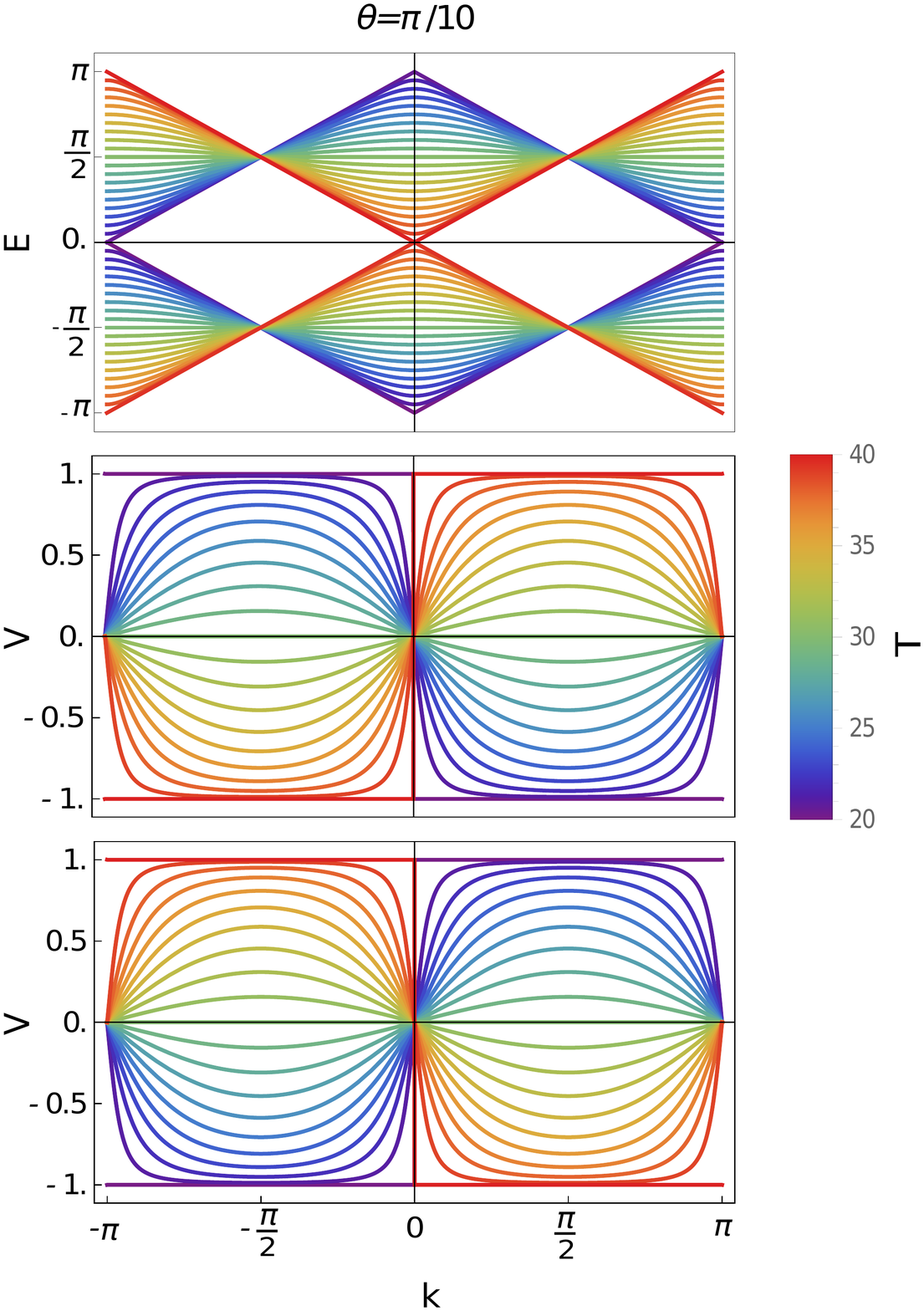}}				
	\end{tabular}}				
	\caption{Energy (up panel) and group velocity (middle panel, positive branch, and down panel, negative branch) as functions of step and momentum for $\theta=\pi/10$. a) For 20 subsequent steps and b) for the subsequent steps of 20 to 40. The energy bands of a specific rotation angle are modified in each step (up panels) and in some steps, the gap between the energies could close up resulting in formation of bound states. Similarly, the group velocity also changes at each step for fixed $\theta$ and through this step dependency, its values changes from positive to negative or vice versa (middle and down panels). This provides the possibility to recognize different topological phases through these (Bloch-oscillating) quantum walks.} \label{Fig1}
\end{figure*}	
\begin{figure}[htb]
	\centering
    {\includegraphics[width=0.85\linewidth]{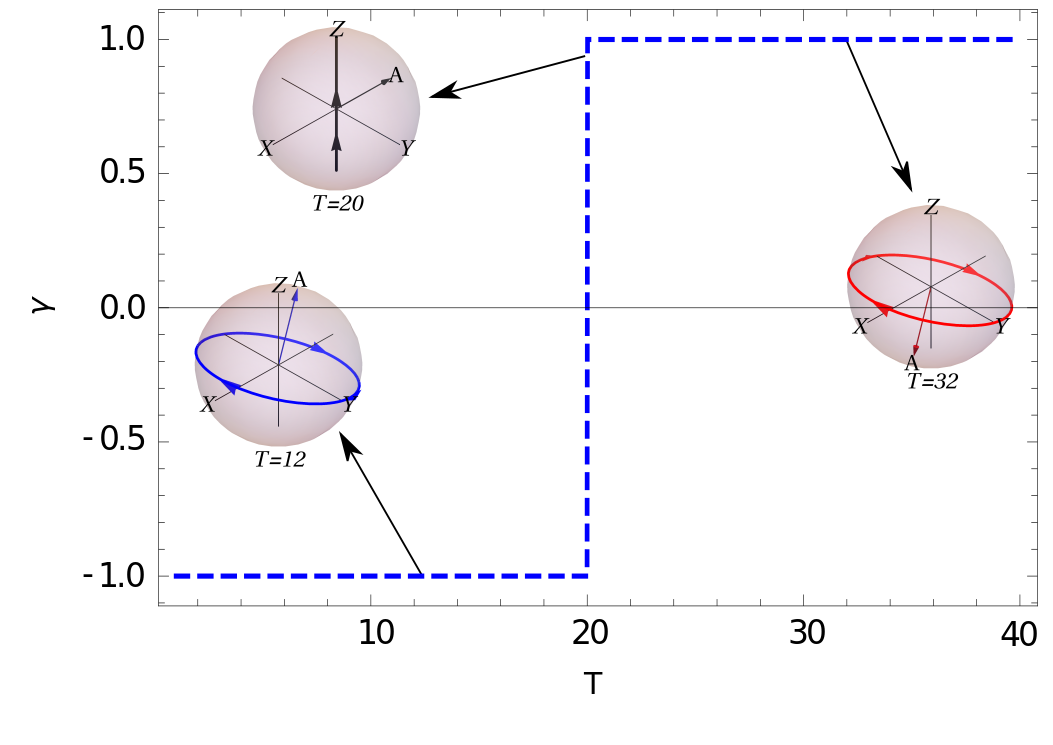}}					
	\caption{Winding number as a function of steps for $\theta=\pi/10$. In the Bloch spheres, the arrows indicate the winding direction of the $\boldsymbol n(k)$ as $k$ traverses through $[-\pi,\pi]$ on a plane perpendicular to vector $\boldsymbol A$. Evidently, for $\gamma=-1$, the winding direction is clockwise while, for $\gamma=1$, it is counterclockwise, confirming two distinguishable phases with different topological invariants.} \label{Fig1.1}
\end{figure}	

\begin{figure*}[htb]
	\centering
	{\begin{tabular}[b]{cc}%
		\subfloat[\label{energyF}]
		{\includegraphics[width=0.33\linewidth]{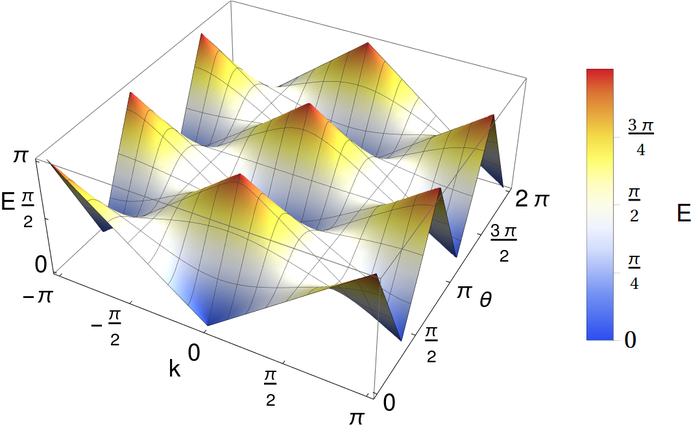}\hspace{1em}
		 \includegraphics[width=0.33\linewidth]{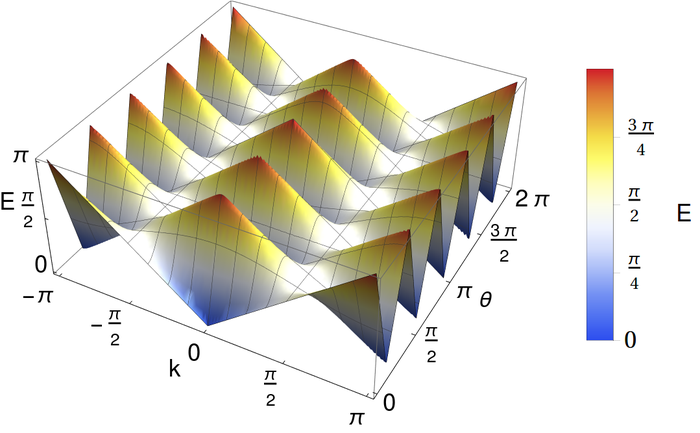}\hspace{1em}
		 \includegraphics[width=0.33\linewidth]{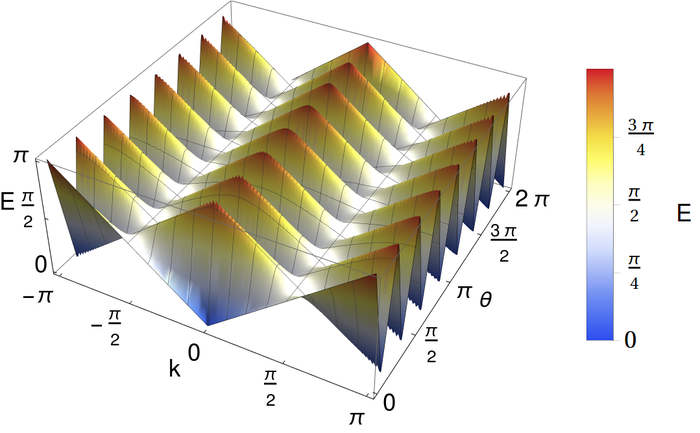}}\\
		\subfloat[\label{groupF}]
		{\includegraphics[width=0.33\linewidth]{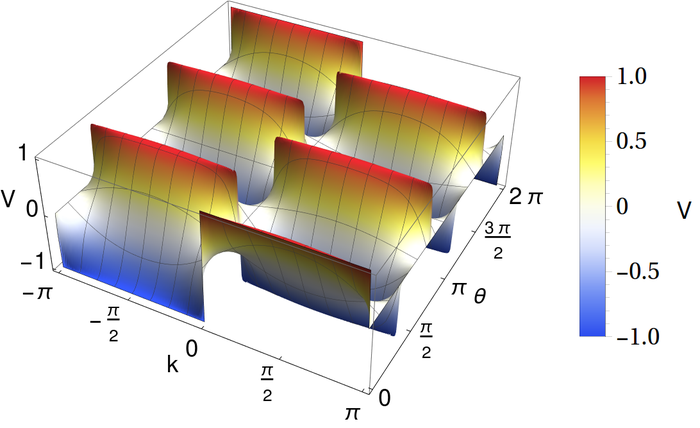}\hspace{1em}
		 \includegraphics[width=0.33\linewidth]{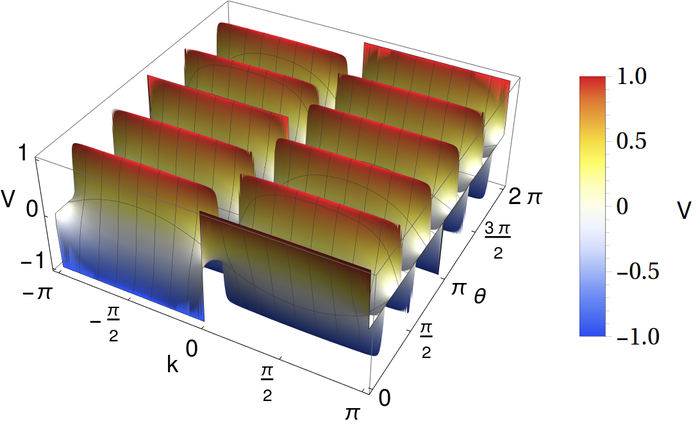}\hspace{1em}
		 \includegraphics[width=0.33\linewidth]{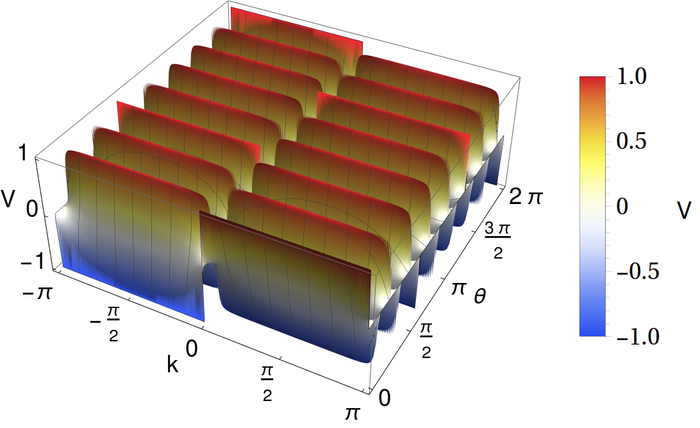}}\\		
		\subfloat[\label{LF}]
	    {\includegraphics[width=0.33\linewidth]{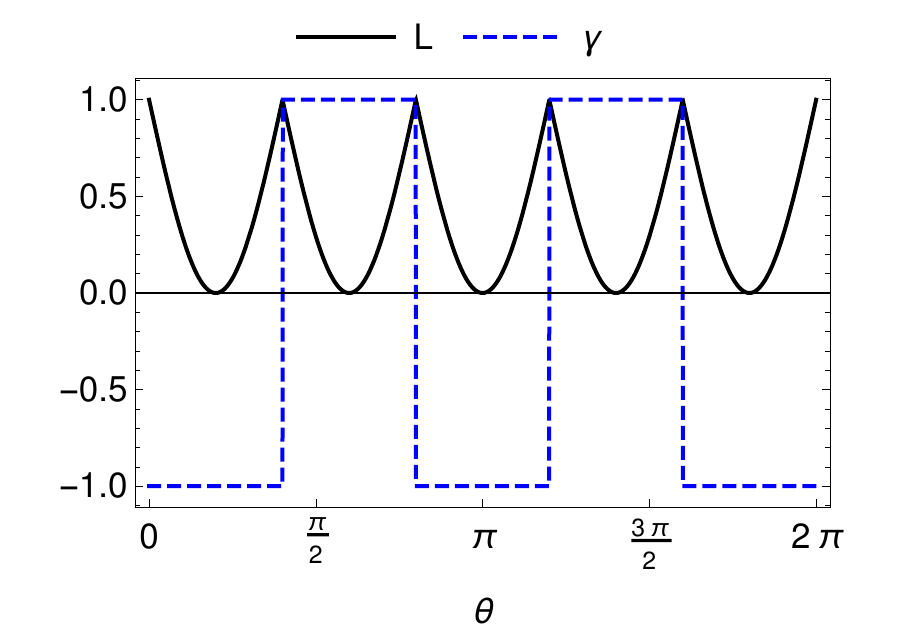}\hspace{1em}
		 \includegraphics[width=0.33\linewidth]{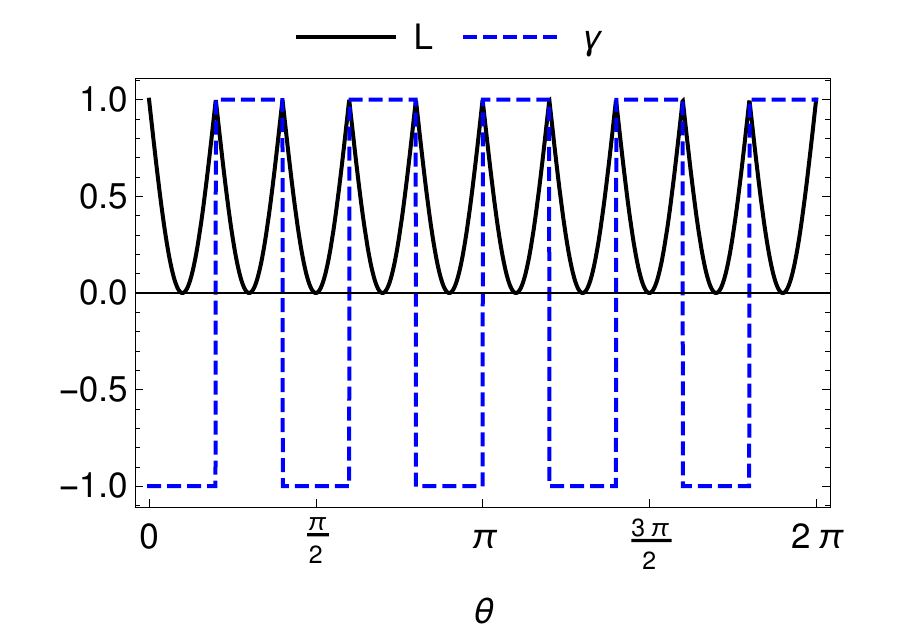}\hspace{1em}
		 \includegraphics[width=0.33\linewidth]{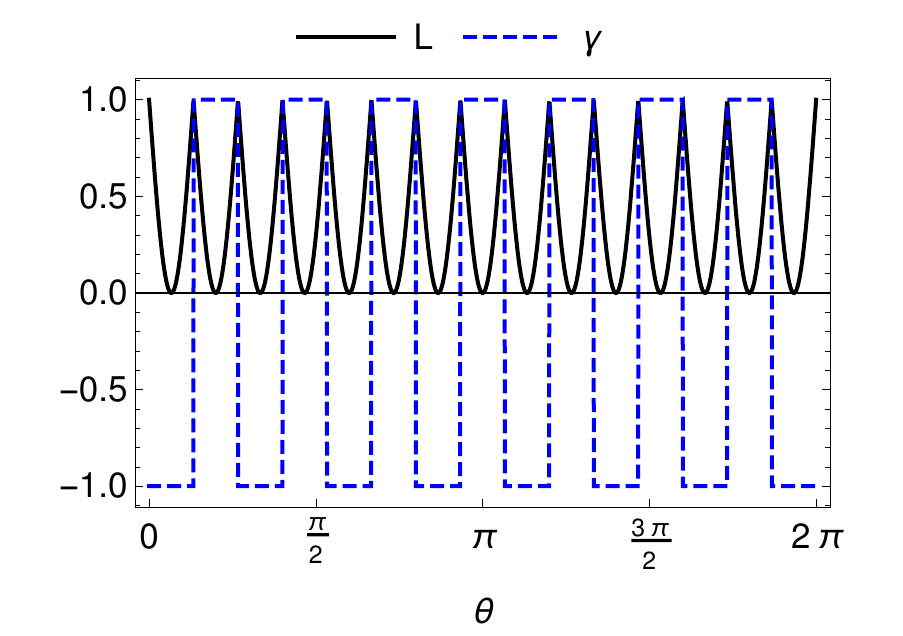}}
    \end{tabular}}				
	\caption{Energy (a), group velocity (b), and $L$ and $\gamma$ (c) as functions of momentum and rotation angle of coin for steps $5$ (vertical left panels), $10$ (vertical middle panels), and $15$ (vertical right panels). We observe that an increment in the number of steps results in the formation of gapless points in energy which indicates the presence of several phases and their corresponding bound states (horizontal up panels). The group velocity (horizontal middle panels) and $L$ (horizontal down panels) are ill-defined in the gapless points and since around each gapless point the winding number ($\gamma$) is different (horizontal down panels), there is a phase transition between two non-trivial phases.} \label{Fig2}
\end{figure*}	
Before we continue, we should point out that momentum, $k$, is limited to the first Brillouin zone since the Hamiltonian has a periodicity given by $H(k+2\pi)=H(k)$. The rotation angle of the coin is also limited to $[0,2\pi]$. All the diagrams are plotted for $k$ and $\theta$ traversing through the first Brillouin zone and $[0,2\pi]$, respectively.

The gap between two bands of energy closes at $E=0$ and $\pm \pi$ (see 
Figs. \ref{Fig1} and \ref{energyF}). Here, the number of times that the energy gap closes is step dependent. For an odd number of steps, the energy gap closes $\frac{T+1}{2}$ times at both $E=0$ and $\pm \pi$, whereas for an even number of steps, the energy gap closes $\frac{T}{2}$ times at $E=0$ and $\frac{T+1}{2}$ times at $\pm \pi$. Therefore, we see the formation of multiple topological phases (multi phase configuration) and their corresponding bound states as quantum walk proceeds. In addition, we notice the parity dependency of the closing gaps at $E=0$ and $\pm \pi$ for odd and even steps. 

The interval ($\theta$ interval) between two gapless bands of energy is a decreasing function of the steps. The first interval between two gapless points, step dependently, modifies as $[0,\frac{2 \pi}{T}]$. This indicates that as the step number increases, the size of the topological phase between two gapless points shrinks and the formations of a large number of topological phases and bound states take place (multi phase configuration). In fact, at each step, there are $T$ phases present. 

In previous studies, it was shown that the bands of energy for specific $\theta$ are not affected by step numbers and remain fixed through evolutions of the quantum walks. In contrast, in Ref. \cite{Ramasesh}, it was shown that if the protocol of the quantum walks is modified to be time dependent (more precisely, position dependent at each step), it will result in a Bloch-oscillating quantum walk which can be used to directly measure topological invariants imprinted as a Berry phase on the quantum state of the particle. The considered protocol for quantum walks in this paper is also Bloch-oscillating \cite{Panahiyan}. The bands of energy associated to each $\theta$ are dynamical and changes from one step to another one (see 
Figs. \ref{G1} and \ref{G2}). This indicates that through stepwise evolutions of quantum walks, the bands of energy for specific $\theta$ could change from one topological phase into another one. Since each phase is characterized by a topological invariants, this step dependency of bands of energy could be used as a method to measure topological invariants, similar to Ref. \cite{Ramasesh}. Later, in our calculations of topological invariants, we address this possibility. 

\subsection{Topological Invariants} \label{TopologyI}

Each topological phase located between two gapless points is characterized by a topological invariant. To specify the topological invariants associated to different phases, one can count the number of times $\boldsymbol n(k)$ winds around the origin (winding number) in a plane orthogonal to $\boldsymbol A$ as $k$ varies throughout the first Brillouin zone. Therefore, the winding number is the topological invariant associated to each phase. The winding number is calculated by \cite{Cardano2017,Asboth2016} 

\begin{eqnarray}
\gamma=\int_{- \pi}^{\pi} \bigg (\boldsymbol n(k) \times \frac{\partial \boldsymbol n(k)}{\partial k} \bigg) \cdot \boldsymbol A \frac{dk}{2 \pi}.
\end{eqnarray} 

In a Hermitian system, the winding number is always an integer \cite{Cardano2017,Asboth2016,Yin2018} while in a non-Hermitian one, it could be half integers \cite{Yin2018,Li2015}. Our coin-shift operator in this paper is unitary and Hermitian; therefore, the resultant winding number is always an integer including $\gamma=0$ and $\pm1$. $\gamma=0$ describes a topologically trivial state while $\gamma=\pm1$ corresponds to non-trivial ones \cite{Yin2018}. 

Our calculations confirms that the winding number takes two values of $\gamma=\pm 1$ (see Figs. \ref{Fig1.1} and \ref{LF}). Positivity or negativity of $\gamma$ depends on the winding direction of the $\boldsymbol n(k)$ as $k$ traverse through $[-\pi,\pi]$. $\gamma=1$ is observed for the counterclockwise winding direction while for $\gamma=-1$, the winding direction is clockwise (see Figs. \ref{Fig1.1}). Therefore, we have two distinguishable topological phases with different topological invariants and any transition from one of them to the other one is through a phase transition. It should be noted that topological phases and their topological invariants simulated by our quantum walks are similar to those observed by the SSH model for polyacetylene (compare  Fig. \ref{CFG} with Fig. \ref{Fig1.1}).

Previously, we showed that there are $T$ phases available at each step and, from one phase to another one, there is a change from one topological invariant to another one. The first phase has winding number of $\gamma=-1$, irrespective of the number of steps. In contrast, the winding number of the final phase is step dependent. For odd steps, the final phase has a winding number of $\gamma=-1$ while for even steps, it is $\gamma=1$ (see Fig. \ref{LF}). Therefore, we have a parity dependency for the number of phases with winding numbers of $\gamma=-1$ and $\gamma=1$ at each step. 

For odd steps, there are $\frac{T+1}{2}$ phases with winding number $\gamma=-1$ and $\frac{T-1}{2}$ phases with winding number $\gamma=1$, whereas, for even number of steps, there are equal numbers of phases with $\gamma=-1$ and $\gamma=1$ given by $\frac{T}{2}$. Interestingly, the phases that start with gapless point $E=0$ and end with gapless point $E=\pm \pi$ have a winding number of $\gamma=-1$ while the phases that start with gapless point $E=\pm \pi$ and end with gapless point $E=0$ have a winding number of $\gamma=1$. This can be used to indirectly measure the topological invariants just by studying the energy. It should be noted that the winding number is ill-defined at gapless points. 

\subsection{Group Velocity and Second Moment} \label{GroupI}

To determine possible phase transitions in the system, one can investigate the group velocity of the energy eigenstates and establish its relation with the moments of the probability density distribution in position space. We find the group velocity as 

\begin{equation}
V(k)= \frac{d E(k)}{dk}= \pm \frac{ \cos(\frac{T \theta}{2})  \sin(k)}{ \sqrt{1- [\cos(\frac{T \theta}{2}) \cos(k)]^2}}. \label{groupv}
\end{equation}

Since there are two bands of energy, there are two group velocities associating to them as well. The group velocity has symmetry of $V(k)=-V(-k)$ and its value span within $[-1,1]$ (see 
Figs. \ref{Fig1} and \ref{groupF}). 

Careful examination of the group velocity confirms that it is related to the vector $\boldsymbol n(k)$ by $n_{z}=-|V(k)|$. This indicates that the places where the gap closes for the energy bands (bound states), $\boldsymbol n(k)$, becomes ill-defined and consequently, group velocity is also ill-defined at a gapless point. While the group velocity is discontinuous at gapless points, it is maximized or minimized around these points and its value is fixed to $V(k)=V=\pm 1$. For a gapless point of $E=0$, the group velocity is $V=-1$ and $V=1$ as $k$ traverse through $[-\pi,0)$ and $(0,\pi]$, respectively. In contrast, for gapless points of $E=\pm \pi$, the opposite, $V=1$ and $V=-1$, is observed for the same regions of $k$. These two distinctive behaviors around different gapless points enable us to determine the type of gapless point and distinguish different phases from one another. 

We can characterize the gapless points, different phases, and phase transitions with the second moment of the probability density distribution defined as $M_{2}=\sum m^{2} P_{m}$ in which $m$ and $P_{m}$ are position and its corresponding probability density \cite{Cardano}. It is a matter of calculation to find a large step-number limit for the walker's distribution moments as \cite{Cardano}

\begin{equation}
M_{2}=\int_{-\pi}^{\pi}  \frac{dk}{2\pi} \bram{\phi_{0}} U^{\dagger}_{T}...U^{\dagger}_{0} (-i)^2 \frac{d^2}{dk^2} U_{0}...U_{T} \ketm{\phi_{0}}, 
\end{equation}
in which $\ketm{\phi_{0}}$ is the coin initial state and one can show that 
\begin{equation}
\frac{M_{2}}{T^2}=L+O(\frac{1}{T^2})=\int_{-\pi}^{\pi}  \frac{dk}{2\pi} [V(k)]^{2}+O(\frac{1}{T^2}). 
\end{equation}	

The first issue is the independency of $M_{2}$ of the coin initial state. The phase transitions happen when $M_{2}$ undergoes abrupt slope variations \cite{Cardano}. It is a matter of calculation to find $L$ as 

\begin{equation}
L=1+ i\sqrt{\frac{\cos (T \theta)-1}{2}}. 
\end{equation}	

The second moment or more precisely, $L$, undergoes abrupt slope variations exactly where gapless points are observed (see Fig. \ref{LF}). These are also places where the winding number changes from $1$ to $-1$ or vise versa and group velocity is ill-defined. Therefore, the gapless points of energy are marking phase-transition points. It is worthwhile to mention that for each phase, irrespective of winding number, $L$ shows similar behavior. This indicates that to have a more comprehensive and detailed picture about phase structure, the second moment of the probability density distribution is necessary but not sufficient.

\section{Conclusion} \label{ConI}

In this paper, we utilized a step dependent coin in the protocol of quantum walks and investigated the topological phases, their topological invariants (winding number), bound states separating these topological phases and possible phase transitions between them.  

We showed that the number of phases and their corresponding bound states are step dependent. An increment in the number of steps results into formation of multiple phases. This is achieved at the cost of reduction in size of each topological phase and decrement in the distance between two gapless points in bands of energy (bound states). The topological invariants for each phase were $\pm1$ showing two different non-trivial phases. At the gapless points, we found that the winding number and group velocity are ill-defined and the second moment of the probability density distribution in position space undergoes an abrupt change. In addition, the winding number changes around these gapless points. These are confirmations, that between two phases with different winding numbers, there is a phase transition taking place over the bound states. Since the number of phases is an increasing function of steps, consequently, the number of phase transitions also increases by steps. Finally, we showed that the behaviors of bands of energy and group velocity around different gapless points in the quantum walks with a step-dependent coin could be used to indirectly measure topological invariants.

The step-dependent coin in quantum walks grants us significant control over the walker's behavior and its properties \cite{Panahiyan,Romanelli,Dhar}. More precisely, the step number is a controlling factor that enables us to determine certain properties of the walker. This, consequently, provides us with a certain level of control over topological phases and bound states that these quantum walks can simulate. For example, on certain level, by changing step number, we can choose where the gapless points (bound states and phase transitions) should be and what type of topological invariants (topological phases) specific points should have. In other words, inclusion of the step dependent coin in the quantum walks results into achieving the controllable quantum systems and controllable simulations by them.   

In the next work, we apply the special protocol of split step quantum walks introduced by Kitagawa \textit{et al.} \cite{Kitagawa} to further enrich the phase space and study the possibility of phenomena such as simulation of Zak phases and the creation of topologically protected bound states. 

\section{Appendix} \label{Append}

The coin-shift operator, $\widehat{U}$, is Hermitian and its determinant is $1$. Therefore, $\widehat{H}$ is traceless resulting into symmetric energy around $E=0$. Eq. \ref{Hamiltonian} indicates that eigenvalues of the $\widehat{U}$ operator ($\lambda$) and energy are related to each other by $\lambda(k)=e^{-i E(k)}$. It is a matter of calculation to find the eigenvalues as 

\begin{eqnarray*}
\lambda(k)= \cos(k) \cos(\frac{T \theta}{2}) \pm \sqrt{[ \cos(k) \cos(\frac{T \theta}{2})]^2-1},
\end{eqnarray*}
where, we find the energy as
\begin{eqnarray*}
\cos (E(k))= \cos(k) \cos(\frac{T \theta}{2}),
\end{eqnarray*}
which, by solving, one can find two bands of energy. We confine $\theta$ to a region of $[0,2 \pi]$ and momentum traverses the first Brillouin zone, $[- \pi,\pi]$. We find that two bands of energy closes their gaps at $E(k)=0$ and $\pm \pi$ with $k=0$. For the sake of completeness, we calculate all the possibilities where energy becomes $E(k)=0$ and $\pi$.

The $E=\pi$ takes place if $\theta$ admits

\begin{eqnarray*}
\theta_{E=\pi}=\frac{\pm 2\cos ^{-1} [-\sec (k)]+4\pi c}{T}=\left\{
\begin{array}{cc} 
\frac{4\pi c}{T} & k=-\pi
\\ [0.2cm]
\frac{\pm 2\pi+4\pi c}{T} & k=0
\\[0.2cm]
\frac{4\pi c}{T} & k=\pi    
\end{array}  
\right.  . 
\end{eqnarray*}
where $c$ is an integer. Similarly, $E=0$ results if the rotation angle is

\begin{eqnarray*}
\theta_{E=0}=\frac{\pm 2\cos ^{-1} [\sec (k)]+4\pi c}{T}=\left\{
\begin{array}{cc} 
\frac{\pm 2\pi+4\pi c}{T} & k=-\pi
\\ [0.2cm]
\frac{4\pi c}{T} & k=0
\\[0.2cm]
\frac{\pm 2\pi+4\pi c}{T} & k=\pi    
\end{array}  
\right.  . 
\end{eqnarray*} 

In our study, we found cases where energy is independent of momentum and it is constant, $E=\frac{\pi}{2}$. This happens step dependently for 

\begin{eqnarray*}
\theta_{E=cte=\pi/2}=\frac {\pm 4 \pi  c + \pi } {T}.
\end{eqnarray*}

One can use Eq. \eqref{Hamiltonian} with obtained energy \eqref{energy} and $\boldsymbol n(k)$ \eqref{n} to find the effective Hamiltonian as 

\begin{equation*}
\widehat{H}= \epsilon \begin{pmatrix}
-\sin(k) \cos(\frac{T \theta}{2}) &  \sin(\frac{T \theta}{2}) [\sin(k)-i \cos(k)] \vspace{0,25cm}\\  
\sin(\frac{T \theta}{2}) [\sin(k)+i \cos(k)] & \sin(k) \cos(\frac{T \theta}{2})
\end{pmatrix},
\end{equation*}
where $\epsilon$ is 

\begin{eqnarray*}
\epsilon=\frac {\cos^{-1}[\cos(k)\cos(\frac{T\theta}{2})]} { \sqrt{1-[ \cos(k) \cos(\frac{T \theta}{2})]^2}}.
\end{eqnarray*}

The $\widehat{\Gamma}$ is given by 

\begin{equation*}
\widehat{\Gamma}= \begin{pmatrix}
 \sin(\frac{T \theta}{2}) &   \cos(\frac{T \theta}{2}) \vspace{0,25cm}\\  
 \cos(\frac{T \theta}{2}) &  -\sin(\frac{T \theta}{2})
\end{pmatrix}.
\end{equation*}

It is a matter of calculation to show that the following relations are valid

\begin{eqnarray*}
\widehat{\Gamma}\widehat{H}\widehat{\Gamma}^{-1}=-\widehat{H}, \\
\widehat{\Gamma}^{-1}\widehat{H}\widehat{\Gamma}=-\widehat{H}, \\
\widehat{\Gamma}\widehat{H}\widehat{\Gamma}=-\widehat{H},
\end{eqnarray*}
and we can find orthogonal sublattice projectors as \cite{Asboth2016}

\begin{eqnarray*}
\widehat{P}_{A}=\frac{1}{2} \bigg(\mathbb{I}+\widehat{\Gamma} \bigg), 
\\ [0.2cm]
\widehat{P}_{B}=\frac{1}{2} \bigg(\mathbb{I}-\widehat{\Gamma} \bigg).
\end{eqnarray*}
in which, $\mathbb{I}$ is the identity operator and they admit the following conditions:

\begin{eqnarray*}
\widehat{P}_{A}\widehat{H}\widehat{P}_{A}=\widehat{P}_{B}\widehat{H}\widehat{P}_{B}=0, 
\\ [0.2cm]
\widehat{P}_{A}+\widehat{P}_{B}=\mathbb{I}, 
\\ [0.2cm]
\widehat{P}_{A}\widehat{P}_{B}=0,
\\ [0.2cm]
\widehat{H}=\widehat{P}_{A}\widehat{H}\widehat{P}_{B}+\widehat{P}_{B}\widehat{H}\widehat{P}_{A}.
\end{eqnarray*}

The group velocity becomes ill-defined at the phase-transition points. Therefore, to find the phase transition points, we calculate roots ($\theta'$) of the denominator of the group velocity \eqref{groupv} as 

\begin{eqnarray*}
\theta'=\left\{
	\begin{array}{cc} 
        \frac{\pm 2\cos ^{-1} [-\sec (k)]+4\pi c}{T}
		\\[0.2cm]
		\frac{\pm 2\cos ^{-1} [\sec (k)]+4\pi c}{T}   
	\end{array}  
	\right.  , 
\end{eqnarray*} 
which are coincidence with points where bands of energy are gapless. Therefore, the closing gaps are where phase transitions take place.

\end{document}